\documentclass[prd,aps,superscriptaddress,showpacs,nofootinbib,eqsecnum,amsfonts,amsmath,
,twocolumn,a4
]{revtex4}
\usepackage{bm,graphics,graphicx,epsfig,color,
}
\begin{document}
\title{Cosmography with the Einstein Telescope}

\author{B.S.~Sathyaprakash}
\affiliation{School of Physics and Astronomy, Cardiff University, 
5, The Parade, Cardiff, UK, CF24 3AA}
\author{B.F.~Schutz$^{1}$}
\affiliation{Max Planck Institute for Gravitational Physics, 
Germany}
\author{C.~Van Den Broeck$^{1}$}
\noaffiliation{}

\begin{abstract}
Einstein Telescope (ET) is a 3rd generation gravitational-wave (GW)
detector that is currently undergoing a design study. ET can detect
millions of compact binary mergers up to redshifts 2-8.  A small
fraction of mergers might be observed in coincidence as gamma-ray
bursts, helping to measure both the luminosity distance and red-shift
to the source.  By fitting these measured values to a cosmological model,
it should be possible to accurately infer the dark energy equation-of-state, 
dark matter and dark energy density parameters.  ET could, therefore, 
herald a new era in cosmology.
\end{abstract}
\maketitle

\setcounter{section}{1}
The goal of modern cosmology is to measure the geometrical and
dynamical properties of the Universe by projecting the observed
parameters onto a cosmological model. The Universe has a lot 
of structure on small scales, but on a scale of about 
100 Mpc 
the distribution of both baryonic (inferred from the
electromagnetic radiation they emit) and  dark matter 
(inferred from large scale streaming motion of galaxies) 
components is quite smooth. It is, therefore, quite
natural to assume that the Universe is homogeneous and
isotropic while describing its large-scale properties.
In such a model, the scale factor $a(t),$ which essentially
gives the proper distance between comoving coordinates, and
curvature of spatial sections $k,$ are the only quantities
that are needed to fully characterize the properties of 
the Universe. The metric of a smooth homogeneous 
and isotropic spacetime is 
\begin{equation}
ds^2 = -dt^2 + a^2(t) \frac{d\sigma^2}{1 - k\sigma^2}
+ \sigma^2 \left (d\theta^2 + \sin^2\theta\, d\varphi^2 \right ),
\nonumber
\end{equation}
where $t$ is the cosmic time-coordinate, $(\sigma,\,\theta,\,
\varphi)$ are the comoving spatial coordinates, and $k$ is
a parameter describing the curvature of the $t=\rm const.$ 
spatial slices. $k=0,\,\pm 1,$ for flat, positively
and negatively curved slices, respectively.
The evolution of $a(t)$ depends on the parameter
$k,$ as well as the ``matter" content of the Universe. The latter
could consist of radiation, baryons, dark matter (DM), dark energy (DE),
and everything else that contributes to the energy-momentum
tensor.  

The Friedman equation, which is one of two Einstein 
equations describing the dynamics of an isotropic and homogeneous
Universe, relates the cosmic scale factor $a(t)$ to the energy
content of the Universe through
\begin{equation}
H(t) = H_0 \left [ \hat\Omega_{\rm M}(t) - \frac{k}{H_0^2a^2} + 
\hat\Omega_\Lambda(t)  \right ]^{1/2},
\label{eq:friedman equation}
\end{equation}
where $H(t)\equiv \dot a(t)/a(t)$ is the Hubble parameter 
($H_0=H(t_P)$ being its value at the present epoch $t_P$), 
while $\hat\Omega_{\rm M}(t)$ and $\hat\Omega_\Lambda(t)$ are the (dimensionless)
energy densities of the DM and DE, respectively.
The above equation has to be supplemented with the 
equation-of-state of DM, assumed to be pressure-less 
fluid $p=0$ [$\hat\Omega_{\rm M}(t)=\Omega_M(1+z)^3,$ where
$\Omega_{\rm M}=\hat\Omega_{\rm M}(t_P)$]
and of DE, assumed to be of the form 
$p = w\rho_{\Lambda}$ [$\hat\Omega_\Lambda(t)=\Omega_\Lambda(1+z)^{3(1+w)},$
where $\Omega_\Lambda=\Omega_\Lambda(t_P)$],
with $w=-1$ corresponding to a cosmological constant.  
The goal of cosmography is to measure $(H_0,\,
\Omega_{\rm M},\, \Omega_\Lambda,\, w,\, k,\, \ldots),$
which essentially determine the large-scale geometry and 
dynamics of the Universe.  In the rest of this paper
we shall assume that the spatial slices are 
flat (i.e., $k=0$).

\begin{figure*}[t]
\vskip -1.0 cm
\includegraphics[width=0.35\textwidth]{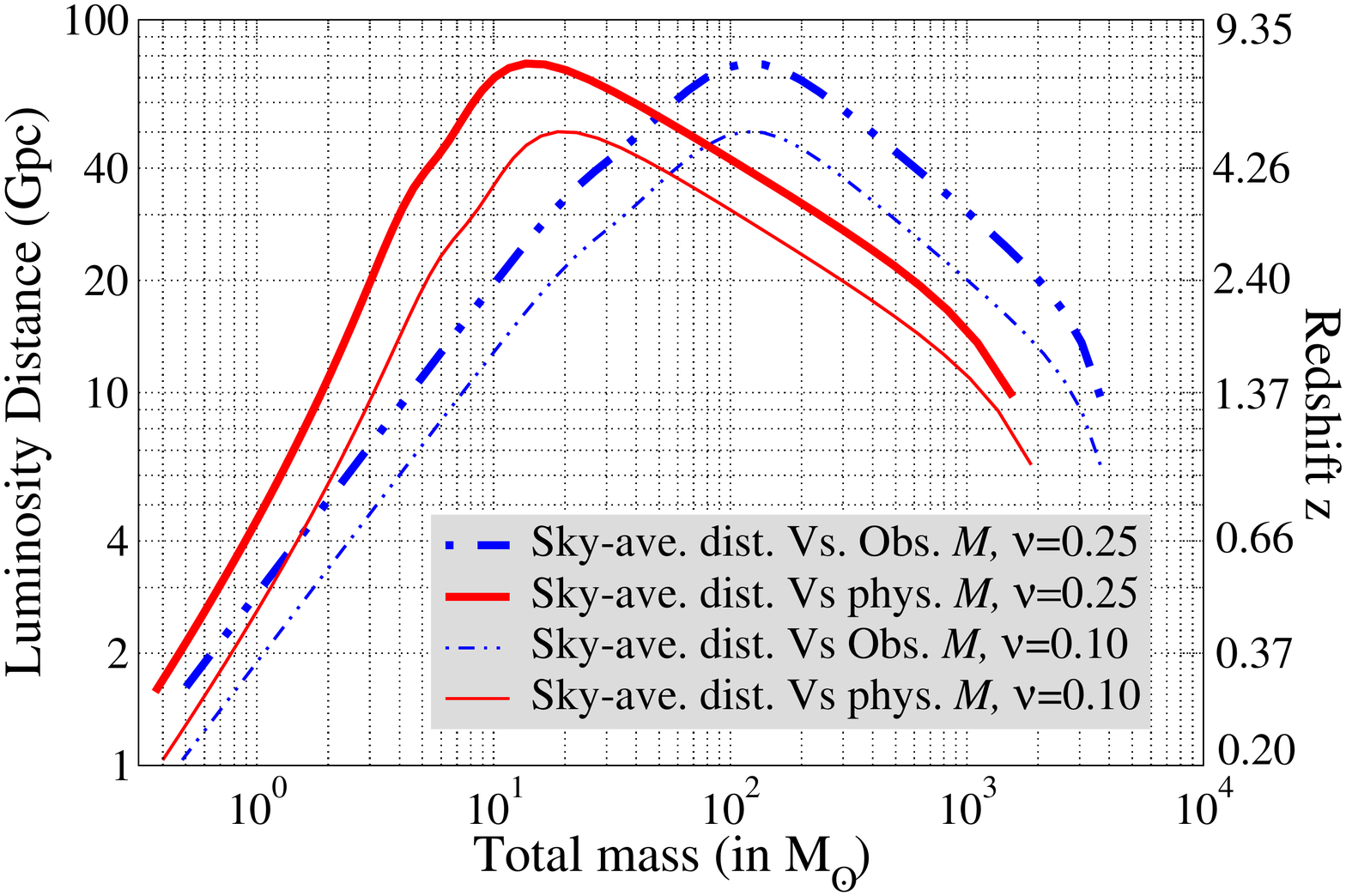}
\includegraphics[width=0.35\textwidth]{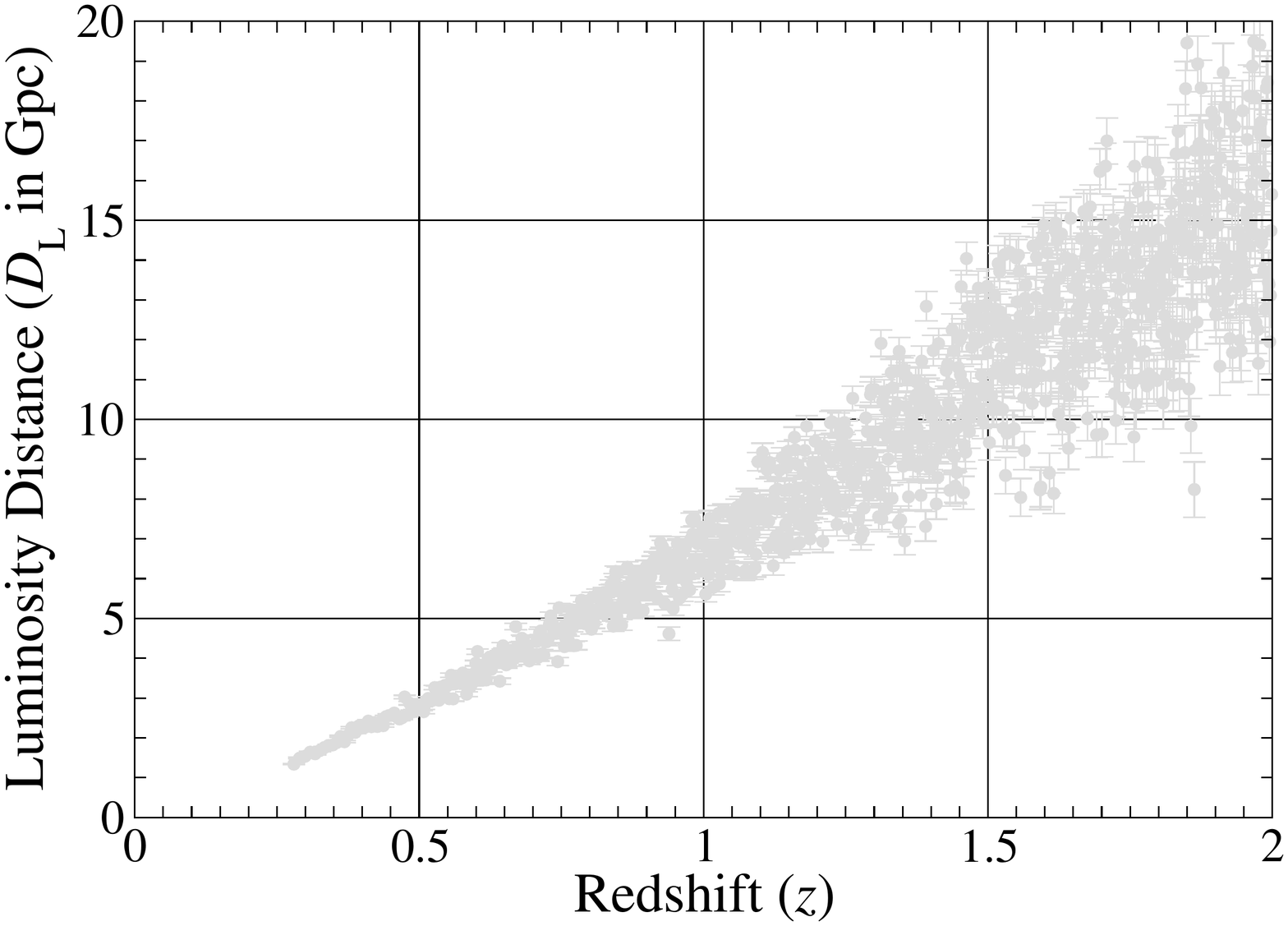}
\vskip -0.5 cm
\caption{The left panel shows the range of the Einstein 
Telescope for inspiral signals from 
binaries as a function of the {\em intrinsic} (red solid line) and 
{\em observed} (blue dashed line) total mass. We assume that a 
source is visible if it produces an SNR of at least 8 in ET.
The right panel shows a realization of the source catalogue showing
the {\em measured} luminosity distance (inferred from GW 
observation of neutron star-black hole mergers) versus their red-shift 
(obtained by optical identification of the source). This catalogue
is then fitted to a cosmological model.}
\label{fig:ET range}
\end{figure*}

Astronomers use ``standard candles'' to measure the geometry of 
the Universe and the various cosmological parameters. A standard 
candle is a source whose intrinsic luminosity $L$ can be inferred 
from the observed properties (such as the spectral content, 
time-variability of the flux of radiation, etc.). Since the 
observations also measure the apparent luminosity $F$, one 
can deduce the luminosity distance $D_{\rm L}$ to a standard
candle from $D_{\rm L}=\sqrt{L/(4\pi F)}.$ 
In addition, if the red-shift $z$ to the source is known then
by observing a population of such sources it will be possible to
measure the various cosmological parameters since the luminosity
distance is related, when $k=0,$ to the red-shift via
\begin{equation}
D_{\rm L} = \frac{c (1+z)}{H_0} \int_0^z \frac{dz'}{\left [ 
\Omega_{\rm M} (1+z')^3 + \Omega_\Lambda (1+z')^{3(1+w)} \right ]^{1/2}}.
\label{eq:cosmology}
\end{equation}
There is no unique standard candle in astronomy that works on all
distance scales. An astronomer, therefore, builds the distance scale 
by using several steps, each of which works over a limited range of
the distance. For instance, the method of parallax can
determine distances to a few kpc, 
Cepheid variables up to $10$ Mpc, the 
Tully-Fisher relation works for several tens of Mpc, the $D_n$-$\sigma$ 
relation up to hundreds of Mpc and Type Ia supernovae up to red-shifts 
of a few \cite{2001ApJ...553...47F}. This way of building the distance scale
has been referred to as the {\em cosmic distance ladder.} For
cosmography, a proper calibration of the distance to high 
red-shift galaxies 
is based on the mutual agreement between different rungs of this 
ladder. It is critical that each of the rungs is calibrated
with as little an error as possible.

Cosmologists have long sought for standard candles that can 
work on large distance scales without being dependent on the 
lower rungs of cosmic distance ladder. In 1986, one of us
pointed out \cite{SCHUTZ1986} that gravitational astronomy can 
provide such a candle, or, more appropriately, a {\em standard 
siren}, in the form of a chirping signal (i.e., a signal whose 
frequency increases as a function of time) from the coalescence 
of compact stars (i.e., neutron stars and black holes) in a binary.
The basic reason for this is that the gravitational-wave (GW)
amplitude depends on the ratio of a certain combination of the
binary masses and the luminosity distance. For chirping signals 
GW observations can measure both the amplitude of the signal and 
the masses very accurately and hence infer the luminosity distance.


Let us first recall in some detail how we might measure the 
luminosity distance. We will first assume that the source
is located close-by, i.e., its redshift $z\ll 1$, although
we will later relax this condition.

Gravitational waves are described by a second rank tensor 
$h_{\alpha\beta}$, which, in a suitable coordinate system and gauge,
has only two independent components $h_+$ and $h_\times,$ $h_{xx}=-h_{yy}=h_+,$
$h_{xy} = h_{yx} = h_\times$, all other components being zero.
A detector measures only a certain linear combination of the
two components, called the response $h(t)$ given by
\begin{equation}
h(t) = F_+(\theta,\, \varphi,\, \psi) h_+(t) +
       F_\times(\theta,\, \varphi,\, \psi) h_\times(t),
\label{eq:response}
\end{equation}
where $F_+$ and $F_\times$ are the detector antenna pattern functions,
$\psi$ is the polarization angle, and $(\theta,\,\varphi)$ are angles
describing the location of the source on the sky. The angles are
all assumed to be constant for a transient source but time-dependent 
for sources that last long enough so that the Doppler modulation 
of the signal due to the relative motion of the source and detector
cannot be neglected. For a coalescing binary consisting 
of two stars of masses $m_1$ and $m_2$ (total mass $M\equiv m_1+m_2$ and 
symmetric mass ratio $\nu\equiv m_1m_2/M^2$) and located at a distance $D_{\rm L}$, 
the GW amplitudes are given by
\begin{eqnarray}
h_+(t) & = & {2\, \nu\, M^{5/3}}\,{D_{\rm L}^{-1}}  (1 + \cos^2(\iota))\,
\omega(t-t_0)^{2/3} \nonumber \\ & \times & \cos [2\Phi(t-t_0;\, M,\,\nu) + \Phi_0'],\nonumber \\
h_\times(t) & = & {4\,\nu\, M^{5/3}}\,{D_{\rm L}^{-1}}  \cos(\iota)\,
\omega(t-t_0)^{2/3} \nonumber \\ & \times & \sin [2\Phi(t-t_0;\, M,\,\nu) + \Phi_0'],
\label{eq:amps}
\end{eqnarray}
where $\iota$ is the angle of inclination of the binary's orbital angular
momentum with the line-of-sight, $\omega(t)$ is the angular velocity 
of the equivalent one-body system around the binary's centre-of-mass and 
$\Phi(t;\, M,\,\nu)$ is the corresponding orbital phase. Parameters $t_0$ and $\Phi_0'$ 
are constants giving the epoch of merger and the orbital phase of the 
binary at that epoch, respectively. 

The above expressions for $h_+$ and $h_\times$ are the dominant terms in 
what is essentially a PN perturbative series.  We have written down the
expressions for a system consisting of non-spinning components on a quasi-circular
orbit. In reality, we cannot assume either to be true. Eccentricity might
be negligible only in the case of stellar mass binaries expected to 
be observed by ground-based detectors, but both eccentricity and spins could 
be non-zero in the case of merger of supermassive black holes expected to
be observed by the Laser Interferometer Space Antenna (LISA) \cite{Apostolatos:1994}.
The argument below holds good to whatever order the amplitudes 
are written down and for non-spinning objects on an eccentric orbit. 
\begin{figure*}[t]
\vskip -0.7 cm
\includegraphics[width=0.4\textwidth]{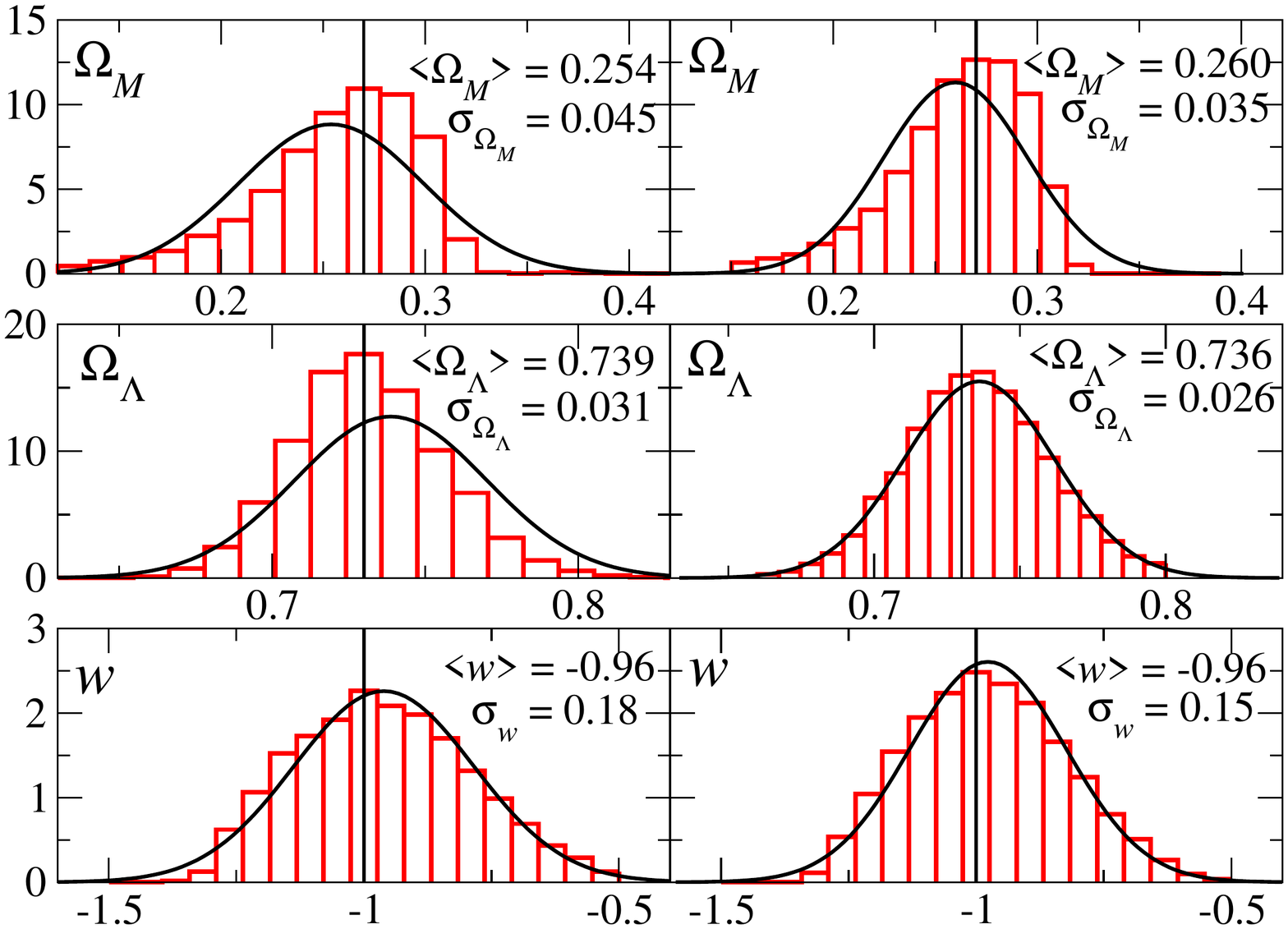}
\includegraphics[width=0.4\textwidth]{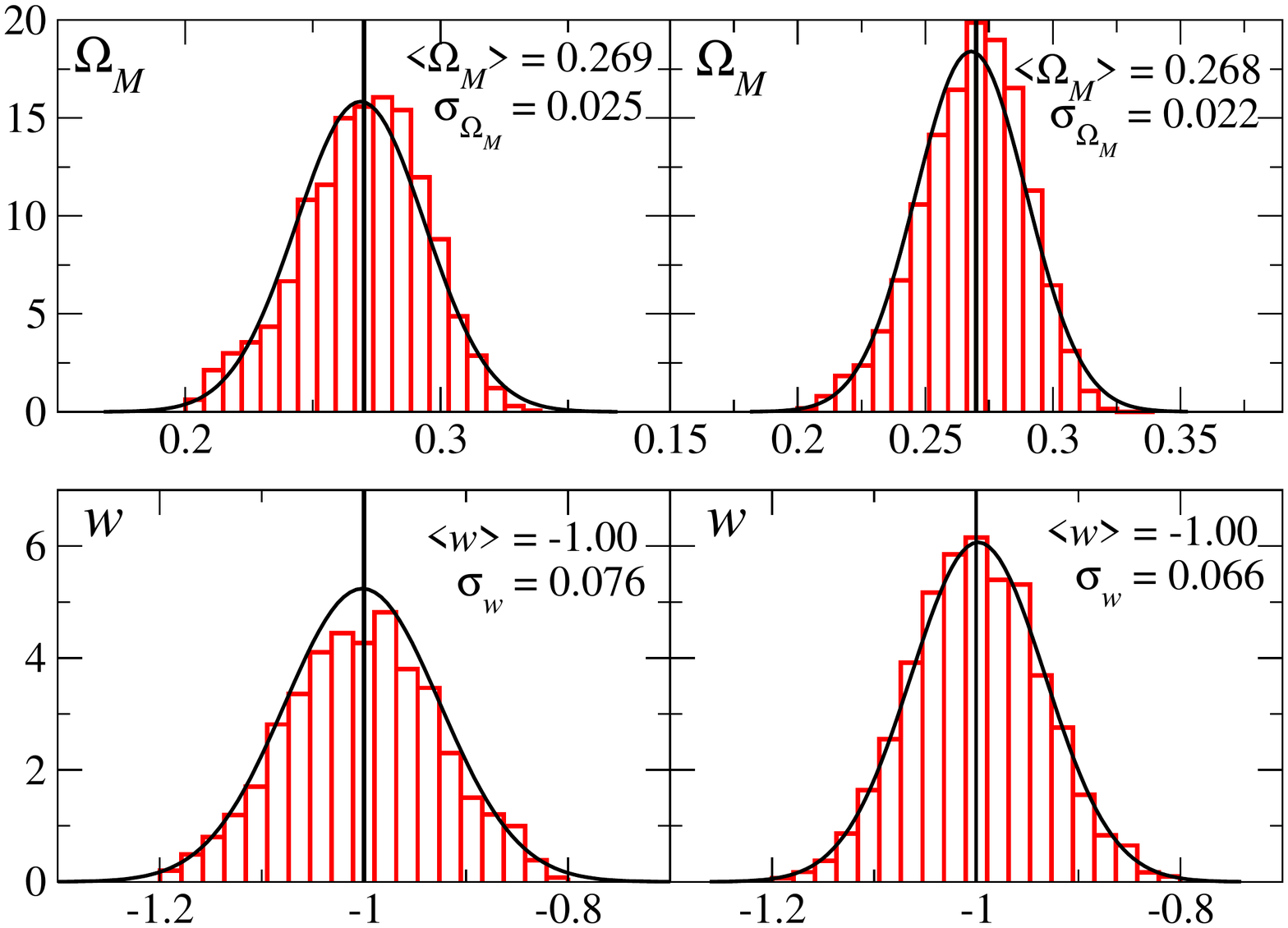}
\vskip -0.5 cm
\caption{The plot on the left shows 
the distribution of errors in $\Omega_{\rm M},$ 
$\Omega_{\Lambda}$ and $w,$ obtained by fitting 5,190 
realizations of a catalogue of BNS merger events to a 
cosmological model of the type given in 
Eq.\,(\ref{eq:cosmology}), with three free parameters.
The fractional 1-$\sigma$ width of the distributions
$\sigma_{\Omega_{\rm M}}/\Omega_{\rm M}$, 
$\sigma_{\Omega_{\Lambda}}/\Omega_{\Lambda}$, and
$\sigma_w/|w|,$ are 18\%, 4.2\% and 18\% 
(with weak lensing errors in $D_{\rm L}$, left panels)
and 14\%, 3.5\% and 15\% (if weak lensing
errors can be corrected, right panels).
The plot on the right is the same, but assuming that 
$\Omega_\Lambda$ is known to be $\Omega_\Lambda=0.73$, and fitting
the ``data'' to the model with two free parameters.  
The fractional 1-$\sigma$ widths in the distribution 
$\sigma_{\Omega_{\rm M}}/\Omega_{\rm M}$ and and $\sigma_w/|w|$,
are 9.4\% and 7.6\% (with weak lensing errors in $D_{\rm L}$, 
left panels) and 8.1\% and 6.6\% (if weak lensing errors 
can be corrected, right panels).}
\label{fig:fits}
\end{figure*}

Substituting the expressions given in Eq.\,(\ref{eq:amps})
for $h_+$ and $h_\times$ in Eq.\,(\ref{eq:response}), we get
\begin{eqnarray}
h(t) & = & \frac{\nu\,M^{5/3}}{D_{\rm eff}}\, \omega^{2/3} \cos[2\Phi(t-t_0;\, M,\,\nu) + \Phi_0],\ \\
D_{\rm eff} & \equiv & \frac{D_{\rm L}}{ \left [ F_+^2 (1+\cos^2(\iota))^2 +
       4 F_\times^2 \cos^2(\iota) \right ]^{1/2}}, \\
\Phi_0 & \equiv & \Phi_0' + \arctan \left [-\frac{2 F_\times\cos(\iota) }{F_+ (1+\cos^2(\iota))}\right ].
\label{eq:response2}
\end{eqnarray}
Here $D_{\rm eff}$ is the effective distance to the binary, which
is a combination of the true luminosity distance and 
the antenna pattern functions. Note that $D_{\rm eff} \ge D_{\rm L}.$
In the case of non-spinning binaries on a quasi-circular orbit,
therefore, the signal is characterized by nine parameters in all, 
$(M, \nu, t_0, \Phi_0, \theta, \varphi, \psi, \iota, D_{\rm L}).$

Since the phase $\Phi(t)$ of the signal is known 
to a high order in PN theory, one employs matched filtering to 
extract the signal and in the process measures the two mass 
parameters $(M,\, \nu)$ (parameters that completely determine 
the phase evolution) and the two fiducial parameters $(t_0,\, \Phi_0).$
In general, the response of a single interferometer will not be 
sufficient to disentangle the luminosity distance from the angular 
parameters. However, EM identification (i.e., electromagnetic, 
especially optical, identification) of the source will determine
the direction to the source, still leaving three unknown parameters
$(\psi,\, \iota,\, D_{\rm L})$. If the signal is a transient, as 
would be the case in ground-based detectors, a network of three 
interferometers will be required to measure all the unknown 
parameters and extract the luminosity distance. 

Although the inspiral signal from a compact binary is a standard 
siren, there is no way of inferring from it the red-shift to a source. 
The mappings $M \rightarrow (1+z) M$, $\omega \rightarrow \omega/(1+z),$ 
and $D_{\rm L} \rightarrow (1+z) D_{\rm L},$ in Eq.\,(\ref{eq:amps}), leave the 
signal invariant.  Note that a source of total mass $M$ at a 
red-shift $z$ will simply appear to an observer to be a binary of 
total mass $(1+z)M$. One must optically identify
the host galaxy to measure its red-shift. Thus, there is
synergy in GW and EM observations which can make precision
cosmography possible, without the need to build a cosmic distance 
ladder.

Over the next two decades GW interferometric detectors will 
provide a new tool for cosmology.  Advanced ground-based 
interferometers, operating around 2015, are 
expected to detect $\sim 40$ binary neutron star mergers 
each year from within about 300 Mpc. Redshift could be 
measured to a (small) number of events associated with GRBs, 
thereby allowing an accurate determination of the Hubble 
constant \cite{Dalal:2006qt,Nissanke:2009kt}. Observation by
LISA of extreme mass ratio inspirals could measure the
Hubble constant pretty accurately \cite{MacLeod:2007jd}.  
LISA will also observe binary super-massive black hole 
mergers with SNRs $\sim \rm few \times 1,000$ enabling 
the measurement of the DE equation-of-state to within 
several percents \cite{HH,Arun:2007hu}.

In the rest of this paper we will discuss how well it might
be possible to constrain cosmological parameters by GW
observations of the inspiral signal of compact binaries
by the Einstein Telescope (ET) --- a third generation GW 
interferometer that is currently under a design study 
\cite{ET}. ET is envisaged to be ten times more sensitive 
than the advanced ground-based detectors, covering a 
frequency range of 1-$10^4$ Hz, posing new challenges in
mitigating gravity gradient, thermal and quantum
noise.  

The sky-position averaged distance up to which ET might detect 
inspiral signals from coalescing binaries with an SNR 
of 8 is shown in Fig.~\ref{fig:ET range}. We plot the range
both as a function of the intrinsic (red solid lines)
and observed (blue dashed lines) total mass. A binary comprising
two $1.4\,M_\odot$-neutron stars (BNS) can be observed
from a red-shift of $z\simeq 2$, and that comprising 
a $1.4\,M_\odot$-neutron star and a $10\,M_\odot$-black hole
(NS-BH) from $z \simeq 4$. 

The expected rate of coalescences per year within the horizon
of ET is $\sim {\rm several}\times 10^5$ for BNS and NS-BH.
Such a large population of events to which
luminosity distances are known pretty accurately,
would be very useful for measuring cosmological parameters.
If, as suspected, BNS and NS-BH are progenitors of short-hard 
gamma-ray bursts (GRBs) \cite{Nakar:2007}, then it might be possible 
to make a coincident detection of a significant subset of
the events in GW and EM windows and obtain
both the luminosity distance to and red-shift of the source.

Since GRBs are believed to be beamed with beaming angles
of order $40^\circ$, we assume 
that only a small fraction ($\sim 10^{-3}$) of binary coalescences 
will have GRB or other EM afterglows that will help us
to locate the source on the sky and measure its
red-shift. Eventually, we will be limited by the 
number of short-hard GRBs observed by detectors that might 
be operating at the time. As a conservative estimate, we 
assume that about $1,000$ BNS mergers will have 
EM counterparts over a three-year period. For definiteness we 
consider only BNS mergers and take
these to have component masses of $(1.4,1.4) M_\odot$.

How well would we measure cosmological parameters with a
catalogue of such sources?
To answer this question we simulated 5,190 realizations
of the catalogue containing 1,000 BNS coalescences
with known red-shift and sky location, but the luminosity
distance subject to statistical errors from GW observation
and weak lensing. One such realization is shown in 
Fig.~\ref{fig:ET range} (right panel).  We assumed that the 
sources were all in the red-shift range $0\le z \le 2$, distributed 
uniformly (i.e., with constant comoving number density) throughout this 
red-shift range. The luminosity distance to the source was computed
by assuming an FRW cosmological model with $H_0=70\,{\rm km\,s^{-1}\,Mpc^{-1}}$,
$\Omega_{\rm M}=0.27$, $\Omega_\Lambda=0.73$, and $w=-1$, but the {\em measured}
distance was drawn from a Gaussian distribution whose width $\sigma_{D_{\rm L}}$
was determined by the quadrature sum of the errors due to weak lensing
and GW observation. Weak lensing error in $D_{\rm L}$
was assumed to be 5\% at $z=1$ and linearly extrapolated to other red-shifts.
GW observational error was estimated from the covariance matrix $C_{km}$ of the 
five-dimensional parameter space of the unknown signal
parameters $p_k = (M,\nu,t_0,\Phi_0, D_{\rm L})$:
\begin{equation}
C_{km} = \Lambda_{km}^{-1},\quad \Lambda_{km} = \left < h_k,\, h_m \right >,
\quad h_k = \frac{\partial h}{\partial p_k}.
\end{equation}
Here the angular brackets denote the scalar product, which,
for any two functions $a(t)$ and $b(t)$, is defined as 
\begin{equation}
\left < a,\, b \right > = 4 \Re \int_0^\infty \frac{{\rm d}f}{S_h(f)}
A(f)\,B^*(f)
\end{equation}
where $A$ and $B$ are the Fourier transforms of the 
functions $a(t)$ and $b(t)$, respectively, and $S_h(f)$ is the ET
noise power spectral density. 
Note that since GRBs are expected to   
be strongly beamed, we did not take the angles $(\iota,\psi)$ associated with the 
unit normal to the plane of the inspiral as unknown variables. This assumption is 
justified: even if the
opening angle of a GRB beam is as large as $40^\circ$, the unit normal
to the plane of the inspiral would still be confined to only 3\% of the area of
a unit sphere. Averaging errors over $(\iota,\psi)$ with the constraint $\iota < 20^\circ$ 
would then be little different from taking $\iota = 0^\circ$. We did, however, average
the errors over the sky position angles $(\theta,\phi)$.
We then fitted each realization of the source catalogue to the 
cosmological model given in Eq.\,(\ref{eq:cosmology}), using the 
Levenberg-Marquardt algorithm \cite{Levenberg:1944,Marquardt:1963}, in order to find a set of best
fit parameters.  It turns out that a catalogue of 1,000 sources is not quite enough 
for an accurate determination of all the parameters.  However, assuming 
that $H_0$ is known accurately, the algorithm gave the best fit
parameters in $(\Omega_{\rm M},\,\Omega_\Lambda,\, w)$ for each of 
the 5,190 realizations. 

The distribution ${\cal P}$ of the parameters obtained
in this way are shown in Fig.\,\ref{fig:fits}, where the 
vertical line is at the true value of the relevant parameter. The relative
1-$\sigma$ errors in $\Omega_{\Lambda},$  $\Omega_{\rm M}$ and $w,$ 
are 4.2\%, 18\% and 18\% (with weak lensing) and
3.5\%, 14\% and 15\% (with weak lensing errors corrected).
Although ${\cal P}(w)$ is quite symmetric, 
${\cal P}(\Omega_{\rm M})$ and ${\cal P}(\Omega_\Lambda)$ 
are both skewed and their mean values are slightly off 
the true values. The medians, however, are coincident with
the true values.  

In addition to $H_0$ if $\Omega_{\Lambda}$ is also known (or, equivalently,
if $\Omega_{\rm M}+\Omega_{\Lambda}=1$), then one can estimate the
pair $(\Omega_{\rm M},\,w)$ more accurately, with the distributions
as shown in Fig.~\ref{fig:fits} with greatly reduced skewness,
and 1-$\sigma$ errors in $\Omega_{\rm M}$ and $w,$ of 9.4\% and 7.6\% 
(with weak lensing) and 8.1\% and 6.6\% (with lensing errors corrected).
Finally, if $w$ is the only parameter unknown, it can be measured to
an even greater accuracy with 1-$\sigma$ errors of 1.4\% (with weak 
lensing) and 1.1\% (with lensing errors corrected).

The results of our simulation are quite encouraging but further
work is needed to confirm the usefulness of GW standard sirens in 
precision cosmology. Let us mention some that are currently being pursued.
Spins of component stars can be legitimately neglected in the case of 
neutron stars (and hence in BNS) but not for black holes. The
modulation in the signal caused by the spin of the black hole 
can improve parameter accuracies. We assumed, for simplicity, that
all our sources are BNS systems with masses $(1.4,\,1.4)M_\odot.$ 
In reality, the catalogue will consist of a range of NS and BH masses. 
A more realistic Monte Carlo simulation would draw binaries from 
the expected population rather than the same system, some of which 
(e.g. more massive systems) would lead to 
better, but others to worsened, parameter accuracies.
The signal contains additional features, such as other harmonics 
of the orbital frequency than the second harmonic considered in this
work and the merger and ringdown signals. These are important
for heavier systems and could potentially reduce the errors. 
These factors are currently being taken into account to
get a more reliable estimation of the usefulness of ET in precision
cosmography.

\bibliography{references}
\end{document}